\documentclass[aps, pra, superscriptaddress, notitlepage, reprint]{revtex4-2}
\usepackage[english]{babel}
\usepackage{amsmath,amsthm}
\usepackage{amsfonts}
\usepackage[pdfborder={0 0 0}, colorlinks=true, urlcolor=blue, linkcolor=blue, citecolor=blue]{hyperref}
\usepackage{color}
\usepackage{graphicx}
\usepackage{float}
\usepackage{subfigure}
\usepackage{lipsum}

\theoremstyle{definition}

\theoremstyle{remark}

\begin{document}

\title{Optomechanical cooling with simultaneous intracavity and extracavity squeezed light}
\author{S. S. Zheng}
\affiliation{Intelligent Science $\&$ Technology Academy of CASIC, Beijing 100041, China}
\affiliation{State Key Laboratory for Mesoscopic Physics, School of Physics, Frontiers Science Center for Nano-optoelectronics, $\&$ Collaborative Innovation Center of Quantum Matter, Peking University, Beijing 100871, China}
\author{F. X. Sun}
\email[Electronic address: ]{sunfengxiao@pku.edu.cn}
\affiliation{State Key Laboratory for Mesoscopic Physics, School of Physics, Frontiers Science Center for Nano-optoelectronics, $\&$ Collaborative Innovation Center of Quantum Matter, Peking University, Beijing 100871, China}
\author{M. Asjad}
\affiliation{Mathematics Department, Khalifa University of Science and Technology, 127788, Abu Dhabi, United Arab Emirates}
\author{G. W. Zhang}
\affiliation{Intelligent Science $\&$ Technology Academy of CASIC, Beijing 100041, China}
\author{J. Huo}
\affiliation{Intelligent Science $\&$ Technology Academy of CASIC, Beijing 100041, China}
\author{J. Li}
\affiliation{Intelligent Science $\&$ Technology Academy of CASIC, Beijing 100041, China}
\author{J. Zhou}
\affiliation{Intelligent Science $\&$ Technology Academy of CASIC, Beijing 100041, China}
\author{Z. Ma}
\affiliation{Intelligent Science $\&$ Technology Academy of CASIC, Beijing 100041, China}
\author{Q. Y. He}
\affiliation{State Key Laboratory for Mesoscopic Physics, School of Physics, Frontiers Science Center for Nano-optoelectronics, $\&$ Collaborative Innovation Center of Quantum Matter, Peking University, Beijing 100871, China}
\affiliation{Collaborative Innovation Center of Extreme Optics, Shanxi University, Taiyuan 030006, China}
\affiliation{Peking University Yangtze Delta Institute of Optoelectronics, Nantong 226010, China}
\affiliation{Hefei National Laboratory, Hefei 230088, China}

\begin{abstract}
We propose a novel and experimentally feasible approach to achieve high-efficiency ground-state cooling of a mechanical oscillator in an optomechanical system under the deeply unresolved sideband condition with the assistance of both intracavity and extracavity squeezing. In the scheme, a degenerate optical parametric amplifier is placed inside the optical cavity, generating the intracavity squeezing; besides, the optical cavity is driven by externally generated squeezing light, namely the extracavity squeezing. The quantum interference effect generated by intracavity squeezing and extracavity squeezing can completely suppress the non-resonant Stokes heating process while greatly enhancing the anti-Stokes cooling process. Therefore, the joint-squeezing scheme is capable of cooling the mechanical oscillators to their quantum ground state in a regime far away from the resolved sideband condition. Compared with other traditional optomechanical cooling schemes, the single-photon cooling rate in this joint-squeezing scheme can be tremendously enlarged by nearly three orders of magnitude. At the same time, the coupling strength required to achieve ground-state cooling can be significantly reduced. This scheme is promising for cooling large-mass and low-frequency mechanical oscillators, which provides a prerequisite for preparing and manipulating non-classical states in macroscopic quantum systems and lays a significant foundation for quantum manipulation.
\end{abstract}
\maketitle

\section{Introduction}
Cavity optomechanics mainly studies the interaction between the electromagnetic radiation field and mechanical degrees of freedom mediated by the radiation pressure force~\cite{Markus2014RMP,PhysicsToday2012,Florian2009Review,MetcalfeMichael2014OptomechanicsReview}. The electromagnetic radiation field ranges from microwave~\cite{Braginsky1967Ponderomotive,Braginsky1970Ponderomotive,Teufel2009MicrowaveMechanicalCoupling} to optical bands~\cite{Dorsel1983PRL_OpticalCavity, Anetsberger2009OpticalMechanicalCoupling,Wilson2015OpticalMechanicalCoupling,Purdy2013OpticalMechanicalCoupling}, meanwhile the vibration frequency and mass of the mechanical oscillator cover a wide range (mass typically ranges from $10^{- 20}$ kilograms to several kilograms, frequency ranges from a few hertz to gigahertz)~\cite{Markus2014RMP,PhysicsToday2012,Florian2009Review,MetcalfeMichael2014OptomechanicsReview}. Owing to these superior characteristics, cavity optomechanical systems have opened up new avenues both in fundamental and applied aspects. From the perspective of fundamental quantum physics, cavity optomechanical systems act as a fascinating platform to study large-scale quantum phenomena~\cite{Ockeloen-Korppi2018Mech-MechEnt,HuangXinyao2018Mech-MechEnt,Thomas2021Mech-MechEnt,Kotler2021Mech-MechEnt,Mercier2021Mech-MechEnt,Clerk2021Mech-MechEnt-Perspective,YuHaocun2020LIGO_QuantumCorrelation}, quantum-classical transitions~\cite{FrancoNori2006QuantumClassicalTransition,TonyELee2013QuantumClassicalTransition} and quantum decoherence~\cite{MarkusAspelmeyer2012OptomechanicsDecoherence}. With regard to applications, cavity optomechanical systems can not only hold great promise for applications in quantum precision measurement (including detecting untrasensitive mass~\cite{Lavrik2003MassDetection,LiuFenfei2013MassDetection}, displacement~\cite{Karabacak2005DisplacementDetection,Schliesser2009Cooling63Phonon_DisplacementDetection,Hoff2013DisplacementDetection_InjectedSqueezing,LiuTianran2020DisplacementSensor}, force~\cite{Moser2013ForceDetection,Fogliano2021ForceSensor}, accelerometer~\cite{OskarPainter2012AccelerometerSensor}, and gravitational waves~\cite{Abbott2016GravitationalWaves,Aasi2013LIGO}); but also pave the way for quantum information processing through implementing optical isolators and circulators~\cite{ShenZhen2016NonreciprocalOptomechanics,ShenZhen2018NonreciprocalOptomechanics}, microwave-optical converter~\cite{WangYingdan2012Transfer,TianLin2012Transfer,Barzanjeh2012Transfer,DongChunhua2012Transfer,Hill2012Transfer,LiuYuxiang2013EIT_Transfer,Andrews2014Transfer,Higginbotham2018Transfer,Bagci2014Transfer}, as well as light storage~\cite{Fiore2011LightStorage,Fiore2013LightStorage}. However, in order to observe quantum effects or prepare non-classical states such as quantum squeezed~\cite{Pirkkalainen2015MechanicalSqueezing,Wollman2015MechanicalSqueezing,Lecocq2015MechanicalSqueezing} or entangled~\cite{Palomaki710,Sun2017,Riedinger2018PCcavity_Ent,ZhengShasha2019PRA,YuHaocun2020LIGO_QuantumCorrelation,Ockeloen-Korppi2018Mech-MechEnt,HuangXinyao2018Mech-MechEnt,Thomas2021Mech-MechEnt,Kotler2021Mech-MechEnt,Mercier2021Mech-MechEnt,Clerk2021Mech-MechEnt-Perspective}
states in macroscopic cavity optomechanical systems serving for certain quantum information tasks, it is essentially necessary to suppress the thermal noise of the surrounding environment firstly, that is, the macroscopic mechanical oscillators need to be cooled to their quantum ground states~\cite{LiuYongChun2013CoolingReview}.    

Theoretical studies have demonstrated that ground-state cooling of mechanical oscillators could be achieved under the resolved sideband condition (that is, the dissipation of the cavity mode is less than the vibration frequency of the mechanical oscillator)~\cite{Marquardt2007CoolingLimit,Wilson-Rae2007CoolingLimit}, which have been experimentally verified by employing various cavity optomechanical systems, such as superconducting microwave cavity~\cite{Teufel2011SDcavity_SBcooling}, photonic crystal microcavity~\cite{Chan2011SB_Cooling}, silicon optomechanical crystal~\cite{LiuQiu2020GroundStateCoolingExperiment}, and others~\cite{SeisYannick2022GroundStateCooling_CavtyElectromechanicalSystem}. However, from a practical perspective, realistic optomechanical systems generally operate in the unresolved sideband regime when the mass of the mechanical oscillator is relatively large and the frequency is relatively low, where the suspended mirror of the laser interferometer gravitational wave detector is a well-known example~\cite{Abbott2016GravitationalWaves,Aasi2013LIGO,Aasi2015_LIGO}.
Therefore, it is necessarily demanded to develop new schemes for ground-state cooling of mechanical oscillators under unresolved sideband conditions. At present, various theoretical and experimental proposals for unresolved sideband regimes have been proposed, including adopting second-order coupling in membrane cavity optomechanical systems to equivalently increase the vibration frequency of the mechanical oscillator~\cite{Deng2012Quadratic_NonSidebandCooling}; utilizing dissipative coupling mechanisms to modulate the dissipation rate of the optical field~\cite{LiMo2009Dissipation_NonSidebandCooling,Clerk2009Dissipation_NonSidebandCooling,Xuereb2011Dissipation_NonSidebandCooling}; adopting measurement-based feedback cooling~\cite{MassimilianoRossi2018GroundStateCooling_unresolved_Feedback}; employing quantum interference effects based on the optomechanically induced transparency phenomenon~\cite{Guo2014EIT_NonSidebandCooling,Ojanen2014EIT_NonSidebandCooling,Liuyongchun2015EIT_NonSidebandCooling,Liuyongchun2015EIT_NonSidebandCooling2,Guwenju2013EIT_NonSidebandCooling} to eliminate the heating effect caused by quantum backaction; and so on. 
Recently, intracavity squeezing~\cite{Clerk2020IntracavitySqueezing_NonSBcooling,GanJinghui2020IntracavitySqueezing_NonSBcooling,Muhhammad2019IntracavitySqueezing_NonSBcooling} or extracavity squeezing ~\cite{Clark2017SqueezingNoise_NonSBcooling,Muhammad2016SqueezingNoise_NonSBcooling} schemes are proposed to achieve ground-state cooling beyond the resolved sideband limit.
Compared with the extracavity squeezing scheme, the intracavity squeezing scheme can achieve better ground-state cooling in a broader range of dissipation. However, it comes with more stringent requirements for parameters. Specifically, an extremely large optical-mechanical coupling strength is necessarily required when the cavity field dissipation is large~\cite{GanJinghui2020IntracavitySqueezing_NonSBcooling}. It is therefore imperative to develop efficient and experiment-friendly ground-state cooling schemes beyond the resolved sideband regime.

%Therefore, novel schemes for ground-state cooling beyond resolved sideband conditions that are both efficient and experiment-friendly need to be further investigated.

In light of the exceptional importance of these previous studies, we propose an innovative and efficient ground-state cooling scheme taking advantage of joint squeezing from both intracavity and extracavity. More specifically, the intracavity squeezing is generated by placing a second-order nonlinear crystal inside the cavity, while the extracavity squeezing results from driving the cavity by externally generated squeezing light. It is worth noting that our proposal is friendly for experimental implementation, as both intracavity~\cite{Furst2011IntracavitySqueezing_WGMcavity_Exp} and extracavity~\cite{Aasi2013LIGO,Hoff2013DisplacementDetection_InjectedSqueezing,Clark2017SqueezingNoise_NonSBcooling} squeezing have already been experimentally demonstrated. 
In this study, we show that by engineering the quantum interference effect generated by intracavity squeezing and extracavity squeezing, the mechanical oscillator in the unresolved sideband condition can be effectively cooled to its quantum ground state. 
Meanwhile, we detailedly analyze the advantages of this joint squeezing scheme compared with conventional sideband cooling, individual intracavity squeezing or individual extracavity squeezing scheme. It is found that our joint squeezing scheme increases the single-photon cooling rate by nearly three orders of magnitude, while greatly reducing the coupling strength required to achieve ground-state cooling.

The structure of this paper is arranged as follows. In Sec.~\ref{Model}, we introduce the theoretical model based on joint squeezing. Then we analyze the expression of the quantum noise spectrum and investigate the condition of completely suppressing the non-resonant Stokes heating process. In Sec.~\ref{NumericalResults}, we numerically simulate the quantum noise spectrum, the effective single-photon cooling rate, as well as the minimum phonon number of the mechanical oscillator that can be achieved. Besides, by comparing our proposal with three conventional ground-state cooling schemes, the advantages of greatly increasing the single-photon cooling rate and reducing the coupling strength are explored. Finally, the content of this paper is summarized in Sec.~\ref{Conclusion}.

\section{Model}\label{Model}

\begin{figure}
\centering
\includegraphics[width=0.9\columnwidth]{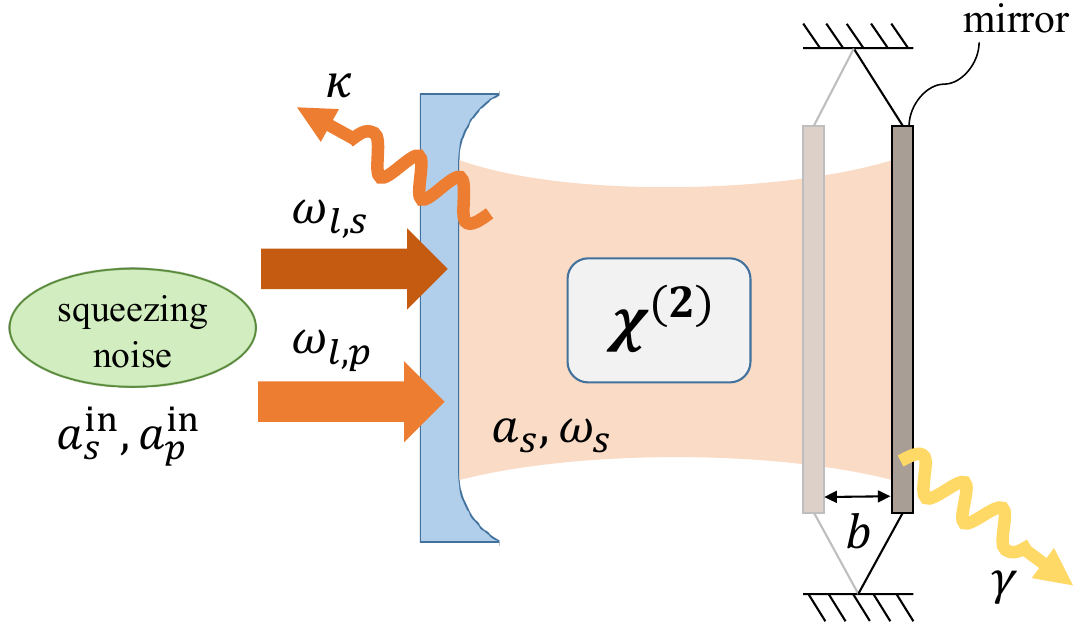}
\caption{A schematic diagram of a standard cavity optomechanical system with both intracavity and extracavity squeezing. A second-order nonlinear crystal is placed inside the cavity to generate intracavity squeezing; meanwhile, an externally generated squeezing light is injected into the cavity from the fixed cavity mirror end, forming the extracavity squeezing noise.}
\label{Cooling_Scheme}
\end{figure}

\subsection{System Hamiltonian and Equations of Motion}
We consider a typical Fabry-P\'erot cavity optomechanical system with the assistance of both intracavity and extracavity squeezing, as displayed in Fig.~\ref{Cooling_Scheme}. On the one hand, a second-order nonlinear crystal is placed inside the cavity to generate degenerate parametric down-conversion, that is, a photon with frequency $\omega_{a_{p}}$ can generate two degenerate photons with frequency $\omega_{a_{s}}=\omega_{a_{p}}/2$, which forms the intracavity squeezing. On the other hand, the externally generated squeezing light drives the cavity from the outside of the fixed mirror, which is the so-called extracavity squeezing. The Hamiltonian of this joint-squeezing scheme can thus be written as ($\hbar=1$)
\begin{eqnarray}\label{Cooling_Hamiltonian_Origin}
H&=&\omega_{a_{s}} a_s^\dagger a_s+\omega_{a_{p}} a_p^\dagger a_p+\omega_m b^\dagger b+(\epsilon_{0} a_s^2a_p^\dagger+\epsilon_{0}^*a_s^{\dagger 2}a_p) \nonumber \\
&&+g_sa_s^\dagger a_s (b^\dagger+b)+g_pa_p^\dagger a_p (b^\dagger+b)+H_{\rm drive}, 
\end{eqnarray}
in which $a_{s}$, $a_{p}$ and $b$ represent the annihilation operators of the optical fundamental mode, the optical pumping mode in the cavity and the mechanical oscillator, respectively. The frequency of the mechanical mode reads $\omega_{m}$. $\epsilon_{0}$ denotes the coupling strength between the optical fundamental mode and the pumping mode, whose quantitative value is related to the second-order nonlinear coefficient $\chi^{(2)}$ of the nonlinear medium. $g_{s}$ ($g_{p}$) describes the corresponding single-photon coupling strength between the intracavity optical fundamental mode (optical pumping mode) and the mechanical mode. And the driving term reads $H_{\rm drive}=E_s e^{i\omega_{l,s} t}a_s+E_p e^{i\omega_{l,p}t}a_p+h.c.$, where $E_{s}$ ($E_{p}$) and $\omega_{l,s}$ ($\omega_{l,p}$) respectively indicate the driving amplitude and the driving frequency of the optical fundamental mode (pumping mode), satisfying the relationship that $\omega_{l,p}=2\omega_{l,s}$. Note that the above Hamiltonian is only associated with the intracavity squeezing term while not explicitly relevant to the extracavity squeezing term, which would be demonstrated later in the equations of motion.

Employing the Heisenberg equations of motion according to the Hamiltonian Eq.~(\ref{Cooling_Hamiltonian_Origin}) and taking account of the dissipation-fluctuation theory, the quantum Langevin equations in the rotating frames relative to the driving laser (i.e., $a_{s}\rightarrow a_{s}e^{-i\omega_{l,s}t}$, $a_{p}\rightarrow a_{p}e^{-i\omega_{l,p}t}$) are obtained as
\begin{eqnarray}\label{Cooling_Langevin0}
\dot{a_{s}} &=&\left(-i \Delta_{s}-\kappa_{s}/2\right) a_{s}-2i\epsilon_{0}^{*} a_{s}^{\dagger} a_{p}-i g_{s} a_{s}\left(b+b^{\dagger}\right) \nonumber \\
&&-E_{s}-\sqrt{\kappa_{s}} a_{s}^{\mathrm{in}}, \nonumber \\
\dot{a_{p}} &=&\left(-i \Delta_{p}-\kappa_{p}/2\right) a_{p}-i\epsilon_{0}a_{s}^{2}-i g_{p} a_{p}\left(b+b^{\dagger}\right) \nonumber \\
&&-E_{p}-\sqrt{\kappa_{p}} a_{p}^{\mathrm{in}}, \nonumber \\
\dot{b} &=&\left(-i \omega_{m}-\gamma/2\right) b-i g_{s} a_{s}^{\dagger} a_{s}-i g_{p} a_{p}^{\dagger} a_{p}-\sqrt{\gamma} b^{\mathrm{in}}.
\end{eqnarray}
$\Delta_{j}=\omega_{j}-\omega_{l,j}$ ($j=s,p$) denotes the detuning between the optical fundamental mode ($j=s$) or the pumping mode ($j=p$) from the corresponding driving laser, respectively. $\kappa_{s}$ ($\kappa_{p}$) represents the total dissipation rate of the optical mode $a_{s}$ ($a_{p}$), while $\gamma$ indicates the corresponding dissipation rate of the mechanical mode. $a_{s}^{\mathrm{in}}$, $a_{p}^{\mathrm{in}}$ and $b^{\rm{in}}$ refer to the relevant input quantum noises arising from the inevitable coupling with surrounding environments.

The quantum noise of the mechanical mode can be characterized by the second-order correlation function $\langle b^{\rm{in}}\left(t\right)b^{\rm{in}\dagger}\left(t'\right)\rangle = (\bar{n}_{\rm{th}}+1)\delta(t-t')$, $\langle b^{\rm{in}\dagger}\left(t\right)b^{\rm{in}}\left(t'\right)\rangle = \bar{n}_{\rm{th}}\delta(t-t')$ with zero mean average, where $\bar{n}_{\rm{th}}=1/[\exp(\hbar\omega_m/k_BT)-1]$ denotes the mean thermal occupation number of the mechanical oscillator, $T$ represents the temperature of the surrounding environment, and $k_B$ describes the Boltzmann constant. Owing to the extracavity squeezing interaction resulting from injecting an externally generated squeezing light into the cavity from the fixed cavity mirror end, the optical fundamental and pumping fields can be regarded as emerging in an squeezed environment~\cite{Clark2017SqueezingNoise_NonSBcooling,Muhammad2016SqueezingNoise_NonSBcooling}, in which the quantum noise of the optical field fulfills the relation $\langle a_{j}^{{\rm in} \dagger}(\omega) a_{j}^{\rm in}(\omega')\rangle=\delta(\omega+\omega')n_s$, $\langle a_{j}^{ \rm in }(\omega) a_{j}^{\rm in}(\omega')\rangle=\delta(\omega+\omega')m_s e^{-2i\phi_s}$ ($j=s,p$), where $n_s=\sinh^2 (r_s),~m_s=\cosh (r_s) \sinh (r_s) =\sqrt{n_s(n_s+1)}$ with $r_s$ being the squeezing factor, and $\phi_s$ represents the relevant squeezing phase of the effective squeezed environment.

When the driving intensity of the laser is relatively strong and the average number of photons in the cavity is relatively large, the standard linearization technique can be performed, in which the field annihilation operator of each mode can be expressed as the sum of the steady-state classical average value and the corresponding quantum fluctuation operator, namely $O=\langle O\rangle+\delta O$ ($O=a_{s},~a_{p},~b,~a_{s}^{\mathrm{in}},~a_{p}^{\mathrm{in}},~b^{\mathrm{in}}$). Substituting them into the Langevin Eq.~(\ref{Cooling_Langevin0}), the classical average value should satisfy the following equations
\begin{eqnarray}
0&=&\left(-i \Delta_{s}-\kappa_{s}/2\right) \alpha_{s}-2i\epsilon_{0}^{*} \alpha_{s}^{*} \alpha_{p}-i g_{s} \alpha_{s}\left(\beta+\beta^{*}\right)-E_{s}, \nonumber \\
0&=&\left(-i \Delta_{p}-\kappa_{p}/2\right) \alpha_{p}-i\epsilon_{0} \alpha_{s}^{2}-i g_{p} \alpha_{p}\left(\beta+\beta^{*}\right)-E_{p}, \nonumber  \\
0&=&\left(-i \omega_{m}-\gamma/2\right) \beta-i g_{s} \alpha_{s}^{*} \alpha_{s}-i g_{p} \alpha_{p}^{*} \alpha_{p},
\end{eqnarray}
in which the average value of each mode is written as $\langle a_{j}\rangle=\alpha_{j}$ ($j=s,p$), and $\langle b\rangle=\beta$, respectively. When the amplitude of the driving laser is large enough to satisfy the condition $|\alpha_{s,p}|\gg1$, the Langevin equation of the quantum fluctuation operators can neglect the high-order terms and only retain the linear terms, therefore the quantum Langevin equations can be formulated as
\begin{eqnarray}\label{Langevin_fluctuation}
 \dot{ \delta a_{s}}&=&\left(-i \Delta_{s}^{\mathrm{ eff }}-\kappa_{s}/2\right) \delta a_{s}-2i\epsilon_{0}^{*} \alpha_{p} \delta a_{s}^{\dagger}-2i\epsilon_{0}^{*} \alpha_{s}^{*} \delta a_{p} \nonumber  \\
&& -i g_{s} \alpha_{s}\left(\delta b+\delta b^{\dagger}\right)-\sqrt{\kappa_{s}} \delta a_{s}^{\mathrm{in}}, \nonumber  \\
 \dot{\delta a _{p}}&=&\left(-i \Delta_{p}^{\mathrm{eff}}-\kappa_{p}/2\right) \delta a_{p}-2i\epsilon_{0} \alpha_{s} \delta a_{s}-\sqrt{\kappa_{p}} \delta a_{p}^{\mathrm{in}} \nonumber  \\
 &&-i g_{p} \alpha_{p}\left(\delta b+\delta b^{\dagger}\right), \nonumber  \\
\dot{\delta b} &=&\left(-i \omega_{m}-\gamma/2\right) \delta b-i g_{s}\left(\alpha_{s}^{*} \delta a_{s}+\alpha_{s} \delta a_{s}^{\dagger}\right) \nonumber  \\
&&-i g_{p}\left(\alpha_{p}^{*} \delta a_{p}+\alpha_{p} \delta a_{p}^{\dagger}\right)-\sqrt{\gamma} \delta b^{\mathrm{in}},
\end{eqnarray}
where the effective detuning reads $\Delta_{j}^{\mathrm{ eff }}=\Delta_{j}+g_{j}(\beta+\beta^{*})$ ($j=s,~p$).

The optical pumping field $a_{p}$ can be adiabatically eliminated when possessing large detuning or dissipation, resulting
\begin{eqnarray}\label{pumping_adiabatic}
\delta a_{p}&=&\frac{1}{i \Delta_{p}^{\mathrm { eff }}+\kappa_p/2}[-2i\epsilon_{0} \alpha_{s} \delta a_{s}-i g_{p} \alpha_{p}\left(\delta b+\delta b^{\dagger}\right) \nonumber \\
&&-\sqrt{\kappa_{p}} \delta a_{p}^{ \mathrm{in}}].
\end{eqnarray}
Inserting this equation into the Langevin equations of motion,  we find that the fundamental shapes of $a_{s}$ and $b$ remain nearly the same, however, the effective detuning, dissipation, coupling strength, and noise terms have a corresponding modification term. The detailed expressions are too sophisticated to be demonstrated here, which will be displayed in Appendix~\ref{HammiltonianDerivation}.
According to Refs.~\cite{GanJinghui2020IntracavitySqueezing_NonSBcooling,Muhammad2016SqueezingNoise_NonSBcooling}, the influence of the pump light field $a_{p}$ on each modification term of $a_{s}$ and $b$ can be ignored or absorbed when $\Delta_{p}^{\mathrm{eff}} \gg \max \left[\sqrt{g_{p}^{2}\left|\alpha_{p}\right|^{2} \kappa_{p} / \omega_{m}}, \sqrt{\kappa_{p} / \kappa_{s}}\left|2\epsilon_{0} \alpha_{s}\right|\right]$. Therefore, the pumping field $a_p$ can be completely regarded as a classical field under the above conditions. We only need to concentrate on the reduced Hamiltonian
\begin{equation}\label{Cooling_EffectiveHamiltonian}
H=\Delta a^\dagger a+\omega_m b^\dagger b+G (a^{\dagger}+a)(b^\dagger+b)+(\epsilon^* a^2+\epsilon a^{\dagger 2}).
\end{equation}
For convenience, the $\delta$ symbols in the quantum fluctuations have been omitted here. Besides, the optical fundamental mode $a_{s}$ is abbreviated as $a$, and the effective detuning is abbreviated as $\Delta=\Delta_{s}^{\mathrm{eff}}$. The effective coupling strength between the optical field and the mechanical oscillator after the linearization procedure denotes $G=g_{s}\alpha_{s}$ (the coupling strength has been set to a real number without loss of generality, which can be easily achieved by adjusting the phase of the driving laser). The last term ($\epsilon^*a^2+\epsilon a^{\dagger 2}$) corresponds to the intracavity squeezing term generated by the second-order nonlinear crystal inside the cavity, where the coupling strength reads $\epsilon=\epsilon_{0}^{*}\alpha_{p}$. We set $\epsilon=|\epsilon|e^{i\phi^\epsilon}$ in the following discussions, in which $|\epsilon|$ and $\phi^\epsilon$ represent the strength and phase of the intracavity squeezing, respectively. According to the reduced Hamiltonian Eq.~(\ref{Cooling_EffectiveHamiltonian}), the Langevin equations of motion for each mode can be simplified as 
\begin{eqnarray}\label{Cooling_Langevin}
\dot{a}&=& -i\Delta a-\frac{\kappa}{2} a-iG(b^\dagger +b)-2i\epsilon a^\dagger -\sqrt{\kappa}a^{\rm in}, \nonumber \\
\dot{b}&=&-i\omega_m b-\frac{\gamma}{2} b -iG(a^{\dagger}+a)-\sqrt{\gamma}b^{\rm in},
\end{eqnarray}
where the optical field is emerged in the squeezing environment (extracavity squeezing), satisfying the relationship $\langle a^{{\rm in} \dagger}(\omega) a^{\rm in}(\omega')\rangle=\delta(\omega+\omega')n_s$, $\langle a^{ \rm in }(\omega) a^{\rm in}(\omega')\rangle=\delta(\omega+\omega')m_s e^{-2i\phi_s}$. The corresponding parameters read $n_s=\sinh^2 (r_s),~m_s=\cosh (r_s) \sinh (r_s) =\sqrt{n_s(n_s+1)}$, where $r_s$ is the external squeezing parameter. It is worth noting that the dissipation $\kappa$ of the optical field generally consists of two parts, which are the external dissipation (the dissipation related to the input-output relationship) $\kappa_{\rm ex}$ and the intrinsic dissipation $\kappa_0$, respectively, i.e., $\kappa=\kappa_{\rm ex}+\kappa_0$. For convenience, we only consider the situation of $\kappa_0=0$ in this work. The case of $\kappa_0\neq0$ is left in the near future.

\subsection{Power spectrum of the radiation pressure force}
In the previous subsection, we introduce the theoretical model and detailly formulize the Langevin equation of motion of the system. In the following, we derive the analytical expression of the quantum noise spectrum of the radiation pressure force, then explore the possibilities whether the present joint-squeezing scheme can completely suppress the quantum backaction in the deeply unresolved sideband regime.

Transforming to the frequency domain, that is, $O(\omega)=\int_{-\infty}^{+\infty} d t e^{i \omega t} O(t)$ ($O=a,b,a^{\dagger},b^{\dagger},a^{\mathrm{in}},b^{\mathrm{in}}$), the Langevin equation (\ref{Cooling_Langevin}) then takes the following form
\begin{eqnarray}\label{Cooling_LangevinFrequency}
-i \omega \tilde{a}(\omega)&=&\left(-i \Delta-\frac{\kappa}{2}\right) \tilde{a}(\omega)-i G\left[\widetilde{b^{\dagger}}(\omega)+\tilde{b}(\omega)\right] \nonumber \\
&&-2i\epsilon \widetilde{a^\dagger}(\omega)-\sqrt{\kappa} \widetilde{a^{\mathrm{in}}}(\omega), \nonumber \\
-i \omega \tilde{b}(\omega)&=&\left(-i \omega_m-\frac{\gamma}{2}\right) \tilde{b}(\omega)-iG\left[ \tilde{a}(\omega)+ \widetilde{a^{\dagger}}(\omega)\right] \nonumber \\
&&-\sqrt{\gamma} \widetilde{b^{\mathrm{in}}}(\omega).
\end{eqnarray}
When the coupling strength between the optical field and the mechanical mode is relatively weak, the perturbation theory can be adopted to theoretically analyze the cooling limit. According to the second formula of Eq.~(\ref{Cooling_LangevinFrequency}), we can get
$\tilde{b}(\omega)\approx \sqrt{\gamma}\widetilde{b^{\mathrm{in}}}(\omega)/\left[i\omega-i\omega_{m}-\gamma/2\right]$. Similarly, by solving the first formula of Eq.~(\ref{Cooling_LangevinFrequency}), the annihilation operator of the optical field can be approximately formalized as 
\begin{eqnarray}
\tilde{a}(\omega)= \frac{\sqrt{\kappa} \chi(\omega) \left[ 2 i \epsilon \chi^{*}(-\omega) \widetilde{a^{\mathrm {in}\dagger}}(\omega)-\widetilde{a^{\mathrm {in}}}(\omega)\right]}{1-4|\epsilon|^{2} \chi(\omega) \chi^{*}(-\omega)},
\end{eqnarray}
in which $\chi(\omega)=1/\left[-i\left(\omega-\Delta\right)+\kappa/2\right]$ represents the optical response function. Therefore, the radiation pressure force acting on the mechanical mode reads 
\begin{eqnarray}
F(\omega)&=&-G\left( \widetilde{a^{\dagger}}(\omega)+ \widetilde{a}(\omega)\right) \nonumber \\
&=&\frac{G \sqrt{\kappa}}{1-4|\varepsilon|^{2} \chi(\omega) \chi^{*}(-\omega)} \bigg\{ \Big[1+2 i \epsilon^{*} \chi^{*}(-\omega) \Big] \chi(\omega) \widetilde{a^{\mathrm{in}}}(\omega) \nonumber \\
&&+\Big[1-2 i \epsilon \chi(\omega)\Big] \chi^{*}(-\omega) \widetilde{a^{\mathrm{in}\dagger}}(\omega) \bigg\}.
\end{eqnarray}
And the power spectrum of the radiation pressure force can be further expressed as 
\begin{eqnarray}\label{Cooling_SpectrumFormula}
S_{\rm FF}(\omega)&=&\int_{-\infty}^{\infty}\langle F(t) F(0)\rangle e^{i \omega t} d t \nonumber \\
&=&S_{\rm FF}^0(\omega)  \bigg|[1+2i\epsilon^*\chi^*(\omega)]\chi(-\omega)\sinh (r_s)e^{-2i\phi_s}  \nonumber \\
&&+[1-2i\epsilon\chi(-\omega)]\chi^*(\omega)\cosh (r_s)\bigg|^2,
\end{eqnarray}
in which $S_{\rm FF}^0(\omega) =G^2\kappa / \big |1-4|\epsilon|^2\chi(\omega)\chi^*(-\omega) \big |^2$. $\omega=\omega_{m}$ corresponds to the anti-Stokes scattering process, which represents the cooling process that the driving laser absorbs a phonon and scatters a photon into the cavity. While $\omega=-\omega_{m}$ corresponds to the Stokes scattering, which describes the heating process that the driving laser simultaneously emits a phonon and a photon in the cavity. We denote the cooling rate and heating rate of the mechanical resonator as $\Gamma_-=S_{\rm FF}(\omega_{m})$ and $\Gamma_+=S_{\rm FF}(-\omega_{m} )$, respectively.

For our joint-squeezing scheme under both extracavity squeezing and intracavity squeezing (abbreviated as ESIS), the cooling and heating rates are
\begin{eqnarray}
\Gamma_-^{\mathrm{ESIS}}&=&S_{\rm FF}(\omega_{m}) \nonumber \\
&=& S_{\rm FF}^0(\omega_m)   \bigg|[1+2i\epsilon^*\chi^*(\omega_{m})]\chi(-\omega_{m})\sinh (r_s)e^{-2i\phi_s} \nonumber \\
&&+[1-2i\epsilon\chi(-\omega_{m})]\chi^*(\omega_{m})\cosh (r_s)\bigg|^2, \nonumber \\
\Gamma_+^{\mathrm{ESIS}}&=&S_{\rm FF}(-\omega_{m}) \nonumber \\
&=&  S_{\rm FF}^0(-\omega_m)  \bigg|[1+2i\epsilon^*\chi^*(-\omega_{m})]\chi(\omega_{m})\sinh (r_s)e^{-2i\phi_s} \nonumber \\
&&+[1-2i\epsilon\chi(\omega_{m})]\chi^*(-\omega_{m})\cosh (r_s)\bigg|^2.
\end{eqnarray}
And the net cooling rate reads $\Gamma_{\mathrm{opt}}^{\mathrm{ESIS}}=\Gamma_{-}^{\mathrm{ESIS}}-\Gamma_{+}^{\mathrm{ESIS}}$. When the system is under the unresolved sideband condition $\kappa>\omega_{m}$, we can set $\Gamma_+^{\mathrm{ESIS}}=S_{\rm FF}(-\omega_m)=0$ for the sake of completely suppressing the heating effect resulted by the Stokes process. Then the condition of completely suppressing the Stokes process can be simplified as 
\begin{equation}
\tanh (r_s) e^{-2i\phi_s}=-\frac{\chi^*(-\omega_m)[1-2i\epsilon \chi(\omega_m)]}{\chi(\omega_m)[1+2i\epsilon^*\chi^*(-\omega_m)]}.
\label{Cooling_Gamma_Stokes}
\end{equation}
And the corresponding anti-Stokes cooling rate takes the following form
\begin{eqnarray}
\Gamma_-^{\mathrm{ESIS}}&=& S_{\rm FF}(\omega_m) \nonumber \\
&=&S_{\rm FF}^0(\omega_m)  \bigg|[1+2i\epsilon^*\chi^*(\omega_m)]\chi(-\omega_m)\sinh (r_s)e^{-2i\phi_s} \nonumber \\
&&+[1-2i\epsilon\chi(-\omega_m)]\chi^*(\omega_m)\cosh (r_s)\bigg|^2.
\label{Cooling_Gamma_AntiStokes}
\end{eqnarray}

In order to analyze the advantages of our proposed extracavity and intracavity joint-squeezing scheme and verify the correctness of the formula (\ref{Cooling_SpectrumFormula}), we select three sets of special parameters to analyze the quantum noise spectrum. They are the common sideband cooling scheme (abbreviated as SB), an independent intracavity squeezing scheme (abbreviated as IS) and an indepentent external squeezing scheme (abbreviated as ES), respectively.

First of all, the joint-squeezing scheme can be reduced to the common sideband cooling scheme when $r_{s}=0,~\epsilon=0$, in which the quantum noise spectrum is simplified as 
\begin{equation}
S_{\rm FF}^{\mathrm{SB}}(\omega)=G^2\kappa\left| \chi(\omega) \right|^{2},
\end{equation}
and the effective cooling and heating rates take the form of
\begin{eqnarray}
\Gamma_-^{\mathrm{SB}}&=&S_{\rm FF}^{\mathrm{SB}}(\omega_{m})=\frac{G^2\kappa}{(\omega_{m}-\Delta)^{2}+\kappa^{2}/4}, \nonumber \\
\Gamma_+^{\mathrm{SB}}&=&S_{\rm FF}^{\mathrm{SB}}(-\omega_{m})=\frac{G^2\kappa}{(\omega_{m}+\Delta)^{2}+\kappa^{2}/4}, \nonumber \\
\Gamma_{\mathrm{opt}}^{\mathrm{SB}}&=&\Gamma_-^{\mathrm{SB}}-\Gamma_+^{\mathrm{SB}} \nonumber \\
&=&\frac{G^2\kappa}{(\omega_{m}-\Delta)^{2}+\kappa^{2}/4}-\frac{G^2\kappa}{(\omega_{m}+\Delta)^{2}+\kappa^{2}/4} .
\end{eqnarray}
These results are in good accordance with Ref.~\cite{Markus2014RMP}.

In the second special case, our proposed scheme is reduced to the scheme where only extracavity squeezing exists when $\epsilon=0$. Then the corresponding quantum noise spectrum reads
\begin{equation}
S_{\rm FF}^{\mathrm{ES}}(\omega)=G^2\kappa \bigg |\chi(-\omega)\sinh (r_s)e^{-2i\phi_s}+\chi^*(\omega)\cosh (r_s)\bigg |^2,
\end{equation}
and the effective cooling and heating rates can be written as
\begin{eqnarray}
\Gamma_-^{\mathrm{ES}}&=&S_{\rm FF}^{\mathrm{ES}}(\omega_{m}) \nonumber \\
&=&G^2\kappa\left| \frac{\sinh (r_s)e^{-2i\phi_s}}{i(\omega_{m}+\Delta)+\kappa/2}+\frac{\cosh (r_s)}{i(\omega_{m}-\Delta)+\kappa/2}\right|^{2}, \nonumber \\
\Gamma_+^{\mathrm{ES}}&=&S_{\rm FF}^{\mathrm{ES}}(-\omega_{m}) \nonumber \\
&=&G^2\kappa\left| \frac{\sinh (r_s)e^{-2i\phi_s}}{i(-\omega_{m}+\Delta)+\kappa/2}+\frac{\cosh (r_s)}{i(-\omega_{m}-\Delta)+\kappa/2}\right|^{2}, \nonumber \\
\end{eqnarray}
which are the same as those in Refs.~\cite{Clark2017SqueezingNoise_NonSBcooling,Muhammad2016SqueezingNoise_NonSBcooling}.

In the third special case, the joint-squeezing proposal is reduced to the case with an independent intracavity squeezing when $r_{s}=0$, in which the quantum noise spectrum becomes
\begin{eqnarray}
S_{\rm FF}^{\mathrm{IS}}(\omega)=\frac{G^2\kappa \bigg | [1-2i\epsilon\chi(-\omega)]\chi^*(\omega)\bigg |^2}{\bigg |1-4|\epsilon|^2\chi(\omega)\chi^*(-\omega) \bigg |^2}.
\end{eqnarray}
The resultant cooling and heating rate read
\begin{eqnarray}
\Gamma_{-}^{\mathrm{IS}}&=&\frac{G^2\kappa \bigg | [1-2i\epsilon\chi(-\omega_{m})]\chi^*(\omega_{m})\bigg |^2}{\bigg |1-4|\epsilon|^2\chi(\omega_{m})\chi^*(-\omega_{m}) \bigg |^2} , \nonumber \\
\Gamma_{+}^{\mathrm{IS}}&=&\frac{G^2\kappa \bigg | [1-2i\epsilon\chi(\omega_{m})]\chi^*(-\omega_{m}) \bigg |^2}{\bigg |1-4|\epsilon|^2\chi(-\omega_{m})\chi^*(\omega_{m}) \bigg |^2}, 
\end{eqnarray}
which are consistent with previous results in Refs.~\cite{Clerk2020IntracavitySqueezing_NonSBcooling,GanJinghui2020IntracavitySqueezing_NonSBcooling,Muhhammad2019IntracavitySqueezing_NonSBcooling}.

\section{Numerical results}\label{NumericalResults}
In the previous section, we derived the Hamiltonian of the system and provided an analytical expression of the quantum noise spectrum. In this section, we will give the numerical simulation results of the quantum noise spectrum of the mechanical oscillator, the effective cooling rate of a single photon, and the minimum phonon number that can be achieved under the unresolved sideband regime. Thus, we are able to compare the advantages of our joint-squeezing scheme with three types of the standard sideband cooling schemes, i.e., the common sideband cooling scheme, the independent intracavity scheme, and the independent extracavity squeezing scheme, respectively.

\subsection{Spectral property and single-photon cooling rate}
In this subsection, we first analyze the quantum noise spectral properties of the mechanical mode and the single-photon cooling rate.

\begin{figure}
\centering
\includegraphics[width=0.9\columnwidth]{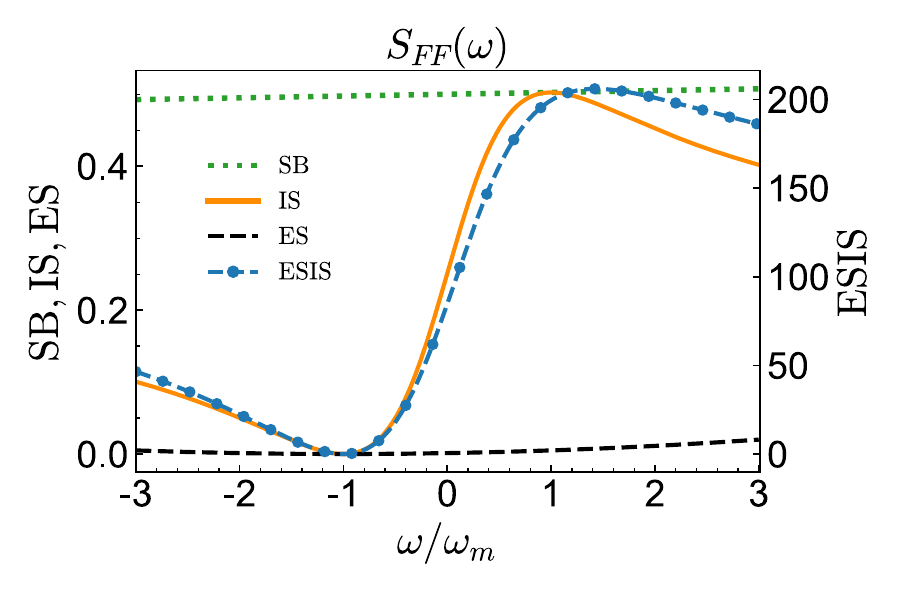}
\caption{The quantum noise spectrum of the four schemes when $\kappa/4\omega_m=100,~\Delta=\sqrt{\omega_m^2+\kappa^2/4}$. The green dotted line represents the standard sideband cooling scheme (SB); the black dotted line denotes the scheme merely with extracavity squeezing (ES); the orange solid line indicates the scheme with independent intracavity squeezing  (IS); while the blue dotted line shows our proposed joint-squeezing scheme by both extracavity and intracavity squeezing (ESIS). The spectrum lines of the SB, ES, and IS regimes use the left axis; while the ESIS scheme adopts the right axis of the dual axis.}
\label{Cooling_Spectrum}
\end{figure}

Figure \ref{Cooling_Spectrum} shows the quantum noise spectrums of four different kinds of cooling schemes when the dissipation of the optical field is large (for example, $\kappa/4\omega_m=100$). It can be clearly found that when the dissipation of the cavity mode $\kappa$ is large, the Stokes process of the standard sideband cooling scheme (SB) can be compared with the anti-Stokes process [i.e. $S_{\mathrm{FF}}^{\mathrm{SB}}(\omega_{m})\sim S_{\mathrm{FF}}^{\mathrm{SB}}(-\omega_{m})$], resulting the small effective cooling rate $\Gamma_{\rm opt}^{\rm SB}=0.005$, as shown by the green dotted line. While the independent extracavity squeezing scheme (ES, shown by the black dashed line) can perfectly suppress the Stokes heating process [i.e., $S_{\mathrm{FF}}^{\mathrm{ES}}(-\omega_{m })=0$], however, the anti-Stokes cooling process will correpondingly be greatly reduced [$S_{\mathrm{FF}}^{\mathrm{ES}}(\omega_{m}) $ is only 0.005]. Therefore, the final effective cooling rate  is as small as $\Gamma_{\rm opt}^{\rm ES}=0.005$, which is approximately the same as the standard sideband cooling scheme (SB). Since the effective single-photon cooling rates of the standard sideband cooling scheme and the independent extracavity squeezing scheme have extremely small values when the cavity field dissipation is large, these two schemes cannot cool the mechanical oscillators to their quantum ground state when the systems are far from the sideband-resolvable conditions. For the independent intracavity squeezing scheme (IS, shown by the solid orange line), it can be clearly seen  that $S_{\mathrm{FF}}^{\mathrm{IS}}(-\omega_{m})= 0$, that is to say, the IS scheme can completely suppress the Stokes heating process, while the anti-Stokes cooling process still maintains a relatively large value, resulting the final effective cooling rate $\Gamma_{\rm opt}^{\rm SB}=0.5$. Therefore, the independent intracavity squeezing scheme can achieve ground-state cooling of mechanical oscillators under indistinguishable sideband conditions. 
For our proposed joint extracavity and intracavity squeezing scheme (ESIS, indicated by the blue dotted line), it can be seen that due to the quantum interference effect between the intracavity squeezing and the extracavity squeezing, not only the Stokes heating process can be completely suppressed [$S_{\mathrm{FF}}^{\mathrm{ESIS}}(-\omega_{m})=0$], but the anti-Stokes cooling process can also be greatly enhanced. Hence, the final effective cooling rate $\Gamma_{\rm opt}^{\rm ESIS}$ is expected to be much larger so that facilitating the ground-state cooling of mechanical oscillators under the deeply unresolved sideband condition.

\begin{figure}
\centering
\includegraphics[width=0.9\columnwidth]{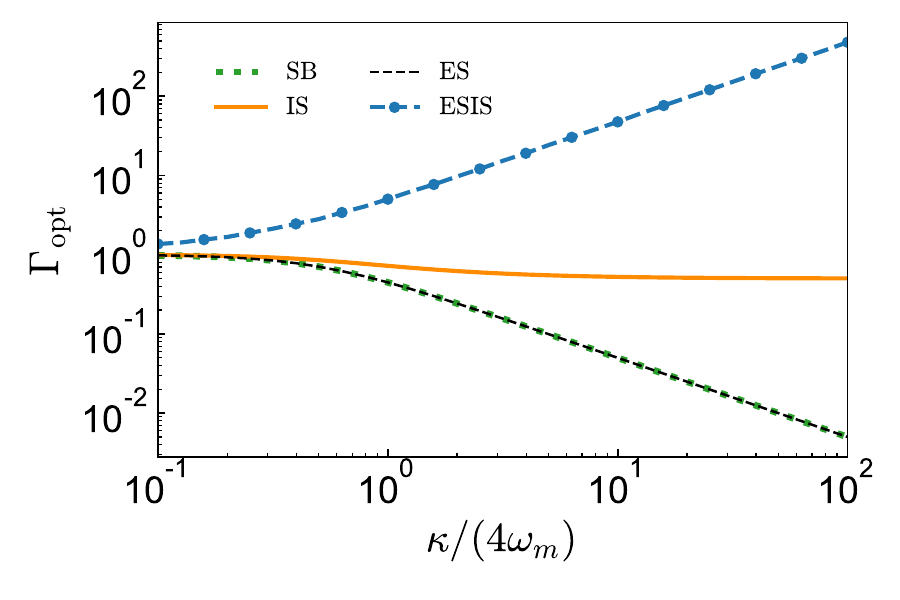}
\caption{The optimal single-photon effective cooling rates of the four schemes vary with the cavity field dissipation $\kappa$ when $\Delta=\sqrt{\omega_m^2+\kappa^2/4}$. Note that the normalization coefficient of $\Gamma_{\rm opt}$ in the figure is $4G^2/\kappa$, which represents the effective single-photon cooling rate. 
}
\label{Cooling_Spectrum_Gamma}
\end{figure}

In order to compare the single-photon cooling rates of the four schemes more clearly, we plot the effective single-photon cooling rates of the four schemes as a function of the dissipation of the cavity field $\kappa$ in Fig.~\ref{Cooling_Spectrum_Gamma}. Note that $\Gamma_{\rm opt}$ is normalized by the coefficient $4G^2/\kappa$. We can easily find that the independent extracavity squeezing scheme (ES, shown by the black dotted curve) and the standard sideband cooling scheme (SB, indicated by the green dotted curve) always have the same effective single-photon cooling rate, which decreases dramatically with the increasing cavity field dissipation and is only $\Gamma_{\rm opt}\sim 10^{-2}$ when the cavity field dissipation increases to $\kappa/(4\omega_{m})=100$. For the single intracavity squeezing scheme (IS, shown by the solid orange curve), the effective single-photon cooling rate decreases slightly as the cavity field dissipation $\kappa$ increases, which still remains about $\Gamma_{\rm opt}^{\rm IS }\approx0.5$ when the cavity field dissipation is rather large. For our proposed extracavity and intracavity joint-squeezing scheme (ESIS, displayed by the blue dotted curve), it is noticed that the effective single-photon cooling rate is slightly larger than those of the other three schemes when the dissipation rate of the cavity is very small (e.g., $\kappa/(4\omega_{m})=10^{-1}$), though the difference is not so obvious. However, it is greatly worth noting that the effective single-photon cooling rate of our joint-squeezing scheme increases exponentially with the cavity field dissipation $\kappa$. In particular, the value of the effective single-photon cooling rate can reach as large as $\Gamma_{\rm opt}^{\rm ESIS}=481$ when the cavity field dissipation is particularly large as $\kappa/(4\omega_{m})=100$, which possesses an improvement of nearly three orders of magnitude compared to the other three proposals. This reflects the remarkable advantage of our proposed intracavity and extracavity joint-squeezing scheme.
\subsection{Minimum phonon number and the required coupling strength}
In the previous subsection, we detailedly analyzed the quantum noise spectrum and single-photon effective cooling rate of the mechanical oscillator, and found that our proposed extracavity and intracavity joint squeezing scheme can greatly increase the single-photon cooling rate by nearly three orders of magnitude comparing with the standard sideband cooling, single intracavity squeezing and single extracavity squeezing schemes. In the following, we will analyze the minimum number of phonons that the mechanical oscillator can be cooled, and discuss the influence of the single-photon cooling rate on the cooling effect.

\begin{figure}
\centering
\includegraphics[width=0.9\columnwidth]{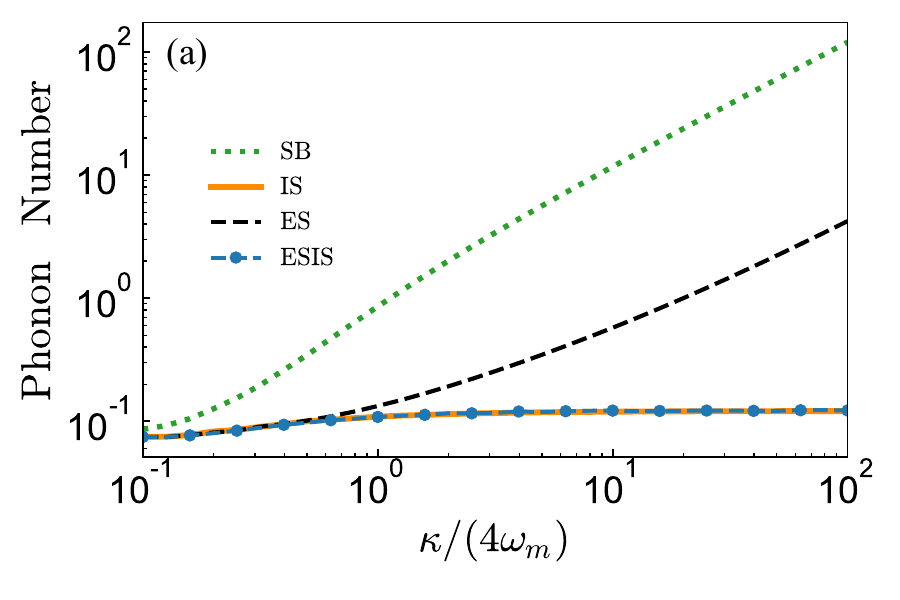}
\includegraphics[width=0.9\columnwidth]{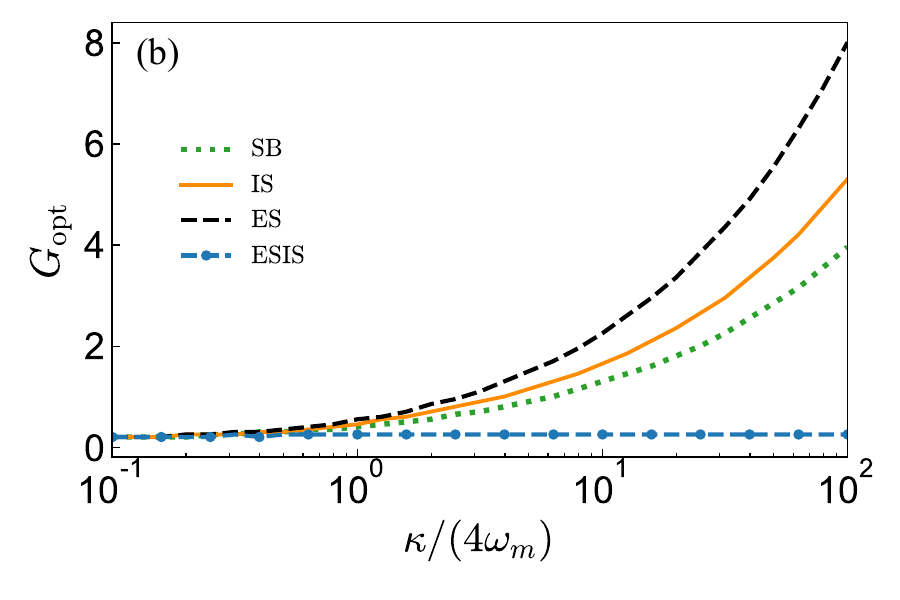}
\caption{(a) Minimum phonon number and (b) the required optical-mechanical coupling strength $G_{\rm opt}$ versus the cavity field dissipation $\kappa$ for the four schemes.
}
\label{Cooling_PhononNumber_G}
\end{figure}

Figure~\ref{Cooling_PhononNumber_G}~(a) shows the variation of the minimum phonon number of the mechanical oscillator with the cavity field dissipation $\kappa$ under the four proposals. It can be easily found that, for the standard sideband cooling scheme (SB, shown by the green dotted line), the optimal phonon number of the mechanical oscillator will be greater than 1 when $\kappa/(4\omega_{m})>1$, meaning that  ground-state cooling cannot be achieved, which is consistent with the theoretical results in the literature \cite{Markus2014RMP}.

For the independent extracavity squeezing scheme (ES), as shown by the black dashed curve in Fig. ~\ref{Cooling_PhononNumber_G}~(a), the minimum steady-state phonon number can be less than 1 when the cavity field dissipation $\kappa/(4\omega_{m})<20$, that is to say, ground-state cooling can be achieved if the cavity field dissipation is not too large. In ref.~\cite{GanJinghui2020IntracavitySqueezing_NonSBcooling}, it is pointed out that the condition for ground-state cooling achieved by an independent extracavity squeezing scheme is $\kappa /\left(4 \omega_{m}\right)<Q_{m} /\left(5 n_{\mathrm {th}}\right)$, where $Q_{m}$ and $n_{\mathrm{th}}$ represent the quality factor and the initial thermal phonon number of the mechanical oscillator, respectively. For the parameters taken in this work $ Q_{m}=10^{5}$, $n_{\mathrm{th}}=10^{3}$, the condition of ground state cooling requires $\kappa /\left(4 \omega_{m}\right) <20$, indicating that our numerical results agree well with this theoretical expression.

For the single intracavity squeezing scheme (IS, indicated by the orange solid line), it can be seen from Fig.~\ref{Cooling_PhononNumber_G}~(a) that the mechanical oscillator can achieve ground-state cooling over a wide range of dissipation parameters, and the minimum phonon number that can be achieved nearly remains constant with the increasing cavity field dissipation $\kappa$. The ref.~\cite{GanJinghui2020IntracavitySqueezing_NonSBcooling} pointed out that the expression of the minimum phonon number denotes $n_{f}^{\mathrm{min} }=2 n_{\mathrm{th}}/Q_{m}+\sqrt{n_{\mathrm{th}}/Q_{m}}$, from which it is clearly seen that the minimum phonon number is irrelevant to the cavity field dissipation $\kappa$.

For our proposed joint-squeezing scheme in the presence of both extracavity and intracavity squeezing (ESIS), as manifested by the blue dash-dotted curve in Fig.~\ref{Cooling_PhononNumber_G}~(a). The final phonon number of the mechanical oscillator is always less than 1 in the domain shown in the diagram. Similar to the independent intracavity squeezing scheme, the minimum phonon number that can be achieved by our joint-squeezing scheme remains essentially unchanged with the cavity field dissipation. Nevertheless, it can be surprisingly observed that although the joint-squeezing scheme can increase the effective single-photon cooling rate by nearly three orders of magnitude, the cooling effect in this regime cannot be effectively improved. The minimum phonon number of the joint-squeezing scheme is much smaller than that of the standard sideband cooling scheme and the independent extracavity squeezing scheme. However, it has almost the same phonon number as the single intracavity squeezing scheme.

Furthermore, we analyze the optical-mechanical coupling strength $G_{\rm opt}$ required to achieve the optimal phonon number, as illustrated in Fig.~\ref{Cooling_PhononNumber_G}~(b). It is easily seen that although the intracavity squeezing scheme (IS, solid orange line) can achieve good ground-state cooling over all the studied dissipative parameters, the coupling strength $G_{\rm opt}$ required to achieve ground state cooling grows almost linearly with increasing dissipative rate $\kappa$. In particular, the minimum number of phonons that can be achieved is $n_{f}^{\rm IS}=0.121$ when $\kappa/(4\omega_{m})=100$, and the required coupling strength corresponds to $G^{\rm IS}_{\rm opt}/\omega_{m}=5.32$. On the other hand, our joint-squeezing scheme although achieves similar cooling effect as the single intracavity squeezing scheme, however, the coupling strength required has been greatly reduced. For example, the optimal phonon number that can be achieved is $n_{f}^{\rm ESIS}=0.1211$ when $\kappa/(4\omega_{m})=100$, and the required coupling strength is only $G^{\rm ESIS}_{\rm opt}/\omega_{m}=0.23$ with the corresponding intracavity squeezing strength $|\epsilon|=141.1$. Furthermore, if we slightly relax the restrictions for achieving ground state cooling, that is, we don't aim at obtaining the minimum phonon number, then the required coupling strength can be largely reduced. When the squeezing strength in the cavity is $|\epsilon|=141.4$, the minimum phonon number obtained after optimization denotes $n_{f}^{\rm IS}=0.4456$. The cooling effect is not optimal in this case, however, the mechanical oscillator can still be cooled to the ground state and the requirement for coupling strength is greatly reduced, which is only $G^{\rm ESIS}_{\rm opt}/\omega_{m}=0.08$. This ensures our joint-squeezing scheme to be experimentally feasible and friendly. Most interestingly, the coupling strength required to achieve the minimum phonon number in the other three schemes (standard sideband cooling regime, independent extracavity squeezing scheme, and independent intracavity squeezing scheme) increases with cavity field dissipation, while the coupling strength $G_{\rm opt}$ required in our proposed joint-squeezing scheme remains nearly unchanged as the cavity field dissipation $\kappa$ increases.

\begin{figure}
\centering
\includegraphics[width=0.9\columnwidth]{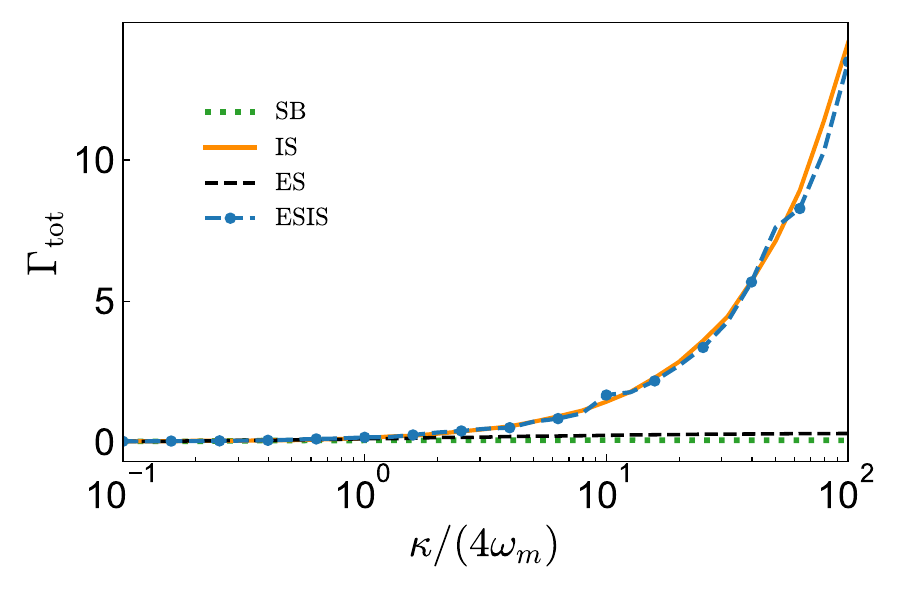}
\caption{Total effective cooling rate $\Gamma_{\rm tot}$ as a function of the cavity field dissipation.
}
\label{Cooling_PhononNumber_GammaG2}
\end{figure}

In brief, our proposed joint squeezing scheme achieves ground-state cooling over an extremely wide range of dissipation parameters. Compared with the other three schemes (standard sideband cooling, independent extracavity squeezing, and independent intracavity squeezing schemes), although the joint-squeezing scheme does not achieve a lower minimum phonon number, it can greatly reduce the requirement for coupling strength to achieve ground-state cooling, which is a significant advantage of this joint-squeezing scheme. This phenomenon can be well explained by Fig.~\ref{Cooling_PhononNumber_GammaG2}, where we plot the variation of the total effective cooling rate $\Gamma_{\rm tot}=\Gamma_{\rm opt}G^2_{\rm opt}$ with the cavity dissipation rate $\kappa$. It can be directly seen from Fig.~\ref{Cooling_PhononNumber_GammaG2} that the total effective cooling rates of both the standard sideband cooling scheme (SB, indicated by the green dotted line) and the independent extracavity squeezing scheme (ES, shown by the black dashed line) possess extremely small values when the cavity dissipation is large. This means that both schemes are inefficient in the regime far away from the sideband resolved condition. Ground-state cooling of the mechanical oscillator in these two schemes is difficult to be achieved. While for the single intracavity squeezing scheme (IS, displayed by the orange solid line) and our proposed theoretical scheme of combined extracavity and intracavity squeezing (ESIS, manifested by the blue dash-dotted line), the total effective cooling rate increases almost exponentially with increasing cavity dissipation $\kappa$. Therefore, it is possible to achieve ground-state cooling in a regime far away from the resolved sideband condition. It is also worth noting that the total effective cooling rates of these two schemes are almost the same, resulting in that the minimum phonon number in the joint-squeezing scheme is almost equal to that of the independent intracavity squeezing scheme. However, since the single photon cooling rate of the joint-squeezing scheme is almost three orders of magnitude larger than that of the independent intracavity squeezing scheme (as has been indicated by Fig. ~\ref{Cooling_Spectrum_Gamma}~), the coupling strength required by the joint-squeezing scheme can be greatly reduced, thus being friendly for experimental realization.

\section{Conclusion}\label{Conclusion}
In summary, we proposed a novel high-efficiency ground-state cooling scheme in the deeply unresolved sideband regime with the assistance of both extracavity and intracavity squeezing. Compared with traditional optomechanical cooling schemes, the single-photon cooling rate can be increased by nearly three orders of magnitude in our proposal. At the same time, the coupling strength required to achieve ground-state cooling can be greatly reduced. Thus, it should be an experimentally feasible scheme to achieve ground-state cooling especially for large-mass and low-frequency mechanical oscillators. This work provides important theoretical support for the study of quantum effects in macroscopic quantum systems and the preparation of nonclassical states in unresolved sideband systems, which plays an important role in promoting quantum manipulation at the macroscopic scale.

\begin{acknowledgements}
We acknowledge illuminating discussions with Y. C. Liu and J. H. Gan.
S. S. Zheng acknowledges the financial support from National Natural Science Foundation of China (Grant No.~12204441). F. X. Sun acknowledges National Natural Science Foundation of China  (Grant No.~12147148) and the China Postdoctoral Science Foundation (Grant No.~2020M680186). M. Asjad has been supported by the Khalifa University of Science and Technology under Award No. FSU-2023-014. Q. Y. He acknowledges National Natural Science Foundation of China (Grant Nos. 11975026, 12125402), and the Innovation Program for Quantum Science and Technology (No.~2021ZD0301500).
\end{acknowledgements}

\appendix\label{APPENDIX}
\section{The derivation of the Hamiltonian (\ref{Cooling_EffectiveHamiltonian})}\label{HammiltonianDerivation}
In the main text, the Langevin equation of the quantum fluctuation operators has been obtained as Eq.~(\ref{Langevin_fluctuation}). Then by  adiabatically eliminating the pumping field in the case of large detuning or dissipation and inserting Eq.~(\ref{pumping_adiabatic}) into Eq.~(\ref{Langevin_fluctuation}), the Langevin equation will take the form of
\begin{widetext}
\begin{eqnarray}\label{Langevin_fluctuation_adiabatic}
 \dot{ \delta a_{s}}&=&\left(-i \Delta_{s}^{\mathrm{ eff }}-\frac{\kappa_{s}}{2}-\frac{4|\epsilon_{0}|^2 |\alpha_{s}|^2}{i \Delta_{p}^{\mathrm { eff }}+\kappa_p/2}\right) \delta a_{s}-2i\epsilon_{0}^{*} \alpha_{p} \delta a_{s}^{\dagger}-i \left(g_{s} \alpha_{s} -i \frac{2\epsilon_{0}^{*} g_{p} \alpha_{s}^{*} \alpha_{p}}{i \Delta_{p}^{\mathrm { eff }}+\kappa_p/2}\right)\left(\delta b+\delta b^{\dagger}\right) \nonumber  \\
&& -\sqrt{\kappa_{s}} \delta a_{s}^{\mathrm{in}}+\frac{2i\epsilon_{0}^{*} \alpha_{s}^{*}}{i \Delta_{p}^{\mathrm { eff }}+\kappa_p/2} \sqrt{\kappa_{p}} \delta a_{p}^{\mathrm{in}}, \nonumber  \\
\dot{\delta b} &=&\left(-i \omega_{m}-\frac{\gamma}{2} +i \frac{2g_{p}^{2}|\alpha_{p}|^2 \Delta_{p}^{\mathrm { eff }}}{ \left(\Delta_{p}^{\mathrm { eff }}\right)^2+\kappa_{p}^{2}/4} \right) \delta b+i \frac{2g_{p}^{2}|\alpha_{p}|^2 \Delta_{p}^{\mathrm { eff }}}{ \left(\Delta_{p}^{\mathrm { eff }}\right)^2+\kappa_{p}^{2}/4}\delta b^{\dagger} -i \left(g_{s}\alpha_{s}^{*}-i \frac{2\epsilon_{0} g_{p} \alpha_{s} \alpha_{p}^{*}}{i \Delta_{p}^{\mathrm { eff }}+\kappa_p/2}\right) \delta a_{s} \nonumber  \\
&&-i \left(g_{s}\alpha_{s}-i \frac{2\epsilon_{0}^{*} g_{p} \alpha_{s}^{*} \alpha_{p}}{i \Delta_{p}^{\mathrm { eff }}-\kappa_p/2}\right) \delta a_{s}^{\dagger}+\frac{i g_{p}\alpha_{p}^{*} }{i \Delta_{p}^{\mathrm { eff }}+\kappa_p/2}\sqrt{\kappa_{p}} \delta a_{p}^{ \mathrm{in}}-\frac{i g_{p}\alpha_{p}}{i \Delta_{p}^{\mathrm { eff }}-\kappa_p/2}\sqrt{\kappa_{p}} \left(\delta a_{p}^{ \mathrm{in}}\right)^{\dagger} -\sqrt{\gamma} \delta b^{\mathrm{in}}.
\end{eqnarray}
\end{widetext}
Therefore, for $\delta a_{s}$, the first term indicates the influence of the optomechanical coupling on the effective detuning $\Delta_{s}^{\mathrm{ eff }} \to \Delta_{s}^{\mathrm{ eff }}-4|\epsilon_{0}|^2 |\alpha_{s}|^2 \Delta_{p}^{\mathrm { eff }}/[(\Delta_{p}^{\mathrm { eff }})^2+\kappa_{p}^{2}/4]$ and the effective dissipation $\kappa_{s} \to \kappa_{s}+4|\epsilon_{0}|^2 |\alpha_{s}|^2 \kappa_{p}/[(\Delta_{p}^{\mathrm { eff }})^2+\kappa_{p}^{2}/4]$. The second term means that the intracavity squeezing effect of the optical fundamental mode is unaffected. The third term suggests that the coupling strength between the optical fundamental mode and the mechanical oscillator is modified. In addition, an additional vacuum noise is also included corresponding to the last term $[2i\epsilon_{0}^{*} \alpha_{s}^{*}\sqrt{\kappa_{p}}/(i \Delta_{p}^{\mathrm { eff }}+\kappa_p/2) ] \delta a_{p}^{\mathrm{in}}$. 

Similarly, for $\delta b$, the first term indicates that the effective detuning is modified $\omega_{m} \to \omega_{m}-2g_{p}^{2}|\alpha_{p}|^2 \Delta_{p}^{\mathrm { eff }}/[(\Delta_{p}^{\mathrm { eff }})^2+\kappa_{p}^{2}/4]$, while the dissipation remains unchanged. The second term suggests that an additional squeezing effect is introduced to the mechanical oscillator. The third and the fourth terms are related to the coupling between the optical fundamental mode and the mechanical oscillator, with the coupling strength modified. And the fifth and the sixth terms correspond to the additional vacuum noises.

According to Refs.~\cite{GanJinghui2020IntracavitySqueezing_NonSBcooling,Muhammad2016SqueezingNoise_NonSBcooling}, the influence of the pump light field $a_{p}$ on each modification term of $a_{s}$ and $b$ can be ignored or absorbed when $\Delta_{p}^{\mathrm{eff}} \gg \max \left[\sqrt{g_{p}^{2}\left|\alpha_{p}\right|^{2} \kappa_{p} / \omega_{m}}, \sqrt{\kappa_{p} / \kappa_{s}}\left|2\epsilon_{0} \alpha_{s}\right|\right]$. In this case, the Langevin equation (\ref{Langevin_fluctuation_adiabatic}) can be further simplified as
\begin{eqnarray}
\dot{ \delta a_{s}}&=&\left(-i \Delta_{s}^{\mathrm{ eff }}-\frac{\kappa_{s}}{2}\right) \delta a_{s}-2i\epsilon_{0}^{*} \alpha_{p} \delta a_{s}^{\dagger}-i g_{s} \alpha_{s} \left(\delta b+\delta b^{\dagger}\right) \nonumber  \\
&& -\sqrt{\kappa_{s}} \delta a_{s}^{\mathrm{in}}, \\
\dot{\delta b} &=&\left(-i \omega_{m}-\frac{\gamma}{2} \right) \delta b -i g_{s}\alpha_{s}^{*} \delta a_{s} -i g_{s}\alpha_{s} \delta a_{s}^{\dagger} -\sqrt{\gamma} \delta b^{\mathrm{in}}. \nonumber  
\end{eqnarray}
In order to make it more clear, we will then change the symbols in the following way: $\delta a_{s} \to a$, $\delta b \to b$, $\Delta_{s}^{\mathrm{ eff }} \to \Delta$, $\epsilon_{0}^{*} \alpha_{p} \to \epsilon$, $g_{s} \alpha_{s} \to G$. And without loss of generality, the effective coupling $G$ can be assumed as real, which can be realized by controlling the initial phase of the driving laser. Thus, we can express the Langevin equation in a more compact form,
\begin{eqnarray}\label{Langevin_fluctuation_simplified}
\dot{a}&=&\left(-i \Delta-{\kappa_{s}}/{2}\right) a-2i\epsilon a^{\dagger}-i G \left(b+b^{\dagger}\right) -\sqrt{\kappa_{s}} a^{\mathrm{in}}, \nonumber \\
\dot{b} &=&\left(-i \omega_{m}-{\gamma}/{2} \right) b -i G (a+a^{\dagger}) -\sqrt{\gamma} b^{\mathrm{in}}.  
\end{eqnarray}
It is directly checked that Eq.~(\ref{Langevin_fluctuation_simplified}) is related to the linearized Hamiltonian, 
\begin{equation}
H=\Delta a^\dagger a+\omega_m b^\dagger b+G (a^{\dagger}+a)(b^\dagger+b)+(\epsilon^* a^2+\epsilon a^{\dagger 2}),
\end{equation}
which is exactly the reduced Hamiltonian (\ref{Cooling_EffectiveHamiltonian}) in the main text.

\bibliography{cooling}
\end{document}